\documentclass[aps,prl,twocolumn]{revtex4-1} 
\usepackage{graphicx}
\usepackage{amssymb}
\usepackage{longtable}
\usepackage{rotating}
\usepackage{braket}

\newcommand{\adr}{De\,R\'{u}jula}

\begin{document}

\title{Examination of the calorimetric spectrum to determine the neutrino mass in low-energy electron capture decay}
\author{R.\,G.\,H.\,Robertson}
\email[]{rghr@uw.edu}
\affiliation{Department of Physics \\
and Center for Experimental Nuclear Physics and Astrophysics, \\
 University of Washington,  Seattle, WA 98195 }
\date{\today}

\begin{abstract}
\begin{description}
\item[Background]  The standard kinematic method for determining neutrino mass from the beta decay of tritium or other isotope is to measure the shape of the electron spectrum near the endpoint.  A similar distortion of the ``visible energy'' remaining after electron capture is caused by neutrino mass.   There has been a resurgence of interest in using this method with $^{163}$Ho, driven by technological advances in microcalorimetry.     Recent theoretical analyses offer reassurance that there are no significant theoretical uncertainties.
\item[Purpose]  The theoretical analyses consider only single vacancy states in the daughter $^{163}$Dy atom.  It is necessary to consider configurations with more than one vacancy that can be populated owing to the change in nuclear charge.
\item[Method]  The shakeup and shakeoff theory of Carlson and Nestor is used as a basis for estimating the population of double-vacancy states.  
\item[Results]  A spectrum of satellites associated with each primary vacancy created by electron capture is presented.   
\item[Conclusions]  The theory of the calorimetric spectrum is more complicated than has been described heretofore.  There are numerous shakeup and shakeoff satellites present across the spectrum, and some may be very near the endpoint.  The spectrum shape is presently not well enough understood to permit a sensitive determination of the neutrino mass in this way.
\end{description}
\end{abstract}

\maketitle

\section{Introduction}

The fact that neutrinos have mass was established by the discovery of neutrino oscillations in atmospheric \cite{Fukuda:1998mi}, solar \cite{Ahmad:2001an}, and reactor \cite{Eguchi:2002dm} neutrinos.   The minimal standard model does not include right-handed fields for neutrinos, and therefore predicts the mass is zero.  How neutrinos acquire their small masses is consequently a matter of great theoretical interest, and may be evidence of new physics at very high mass scales.  Oscillation data provide only the differences between the squares of masses, but do constrain the average mass of the 3 species to be at least 0.02 eV because no squared mass can be less than zero.  Laboratory measurements of the beta spectrum of tritium \cite{Kraus:2004zw,Aseev:2011dq} yield an upper limit on the absolute scale of neutrino mass of less than 2 eV.   Given that the mass must then lie in this range, new, sensitive laboratory measurements are being pursued \cite{Drexlin:2013lha,Monreal:2009za,Andreotti:2007eq} to shed further light on the mechanism for neutrino mass generation.

Neutrinos are also an abundant ingredient of the universe, created in numbers comparable to photons during the big bang.  The combination of direct laboratory measurements and neutrino oscillation data shows that neutrino mass is too small for active neutrinos to be the dark matter that makes up some 27\% of the energy density of the universe, but their mass may influence large-scale structure and other observables.  A laboratory measurement of the mass at an improved level of sensitivity would be valuable in helping to constrain cosmological parameters that are correlated with it, such as the equation of state of dark energy and the fluctuation amplitude of the matter power spectrum \cite{Palanque-Delabrouille:2014jca}.  

Among the ideas being investigated for a laboratory measurement of neutrino mass is one originally proposed more than 30 years ago  \cite{DeRujula:1981ti,DeRujula1982429}, a measurement of the energy retained following electron capture in $^{163}$Ho, a nucleus with a particularly low Q-value \cite{Kopp:2009yp} for the decay to the ground state of $^{163}$Dy.    In this note we raise a concern that, technological progress notwithstanding,  the theoretical description of the spectrum is insufficiently  understood yet to permit an eV-scale determination of the neutrino mass.

\section{Electron-capture decay}

In its simplest form,  electron-capture decay is the capture by the nucleus of a bound atomic electron with the release of an electron neutrino.  The neutrino's energy is the Q-value minus the electron binding energy, and thus consists of several mononergetic lines.   
\begin{eqnarray}
^AZ&\rightarrow {^A(Z-1)}_i+\nu_e+Q_i
\end{eqnarray}
where $A$ and $Z$ are an atomic mass and number, respectively, and $Q_i$ refers to the Q-value for the particular atomic final state $i$.  In this form there is very little sensitivity to neutrino mass because the neutrino is always relativistic.   However, in the early 1980s \adr\ and Lusignoli  \cite{DeRujula:1981ti,DeRujula1982429} recognized that the lines are in fact not monoenergetic because atomic vacancies have  short lifetimes and therefore non-negligible widths.  The decay process is then formally the same as a radiative decay, 
\begin{eqnarray}
^AZ&\rightarrow {^A(Z-1)}+\nu_e+\gamma_i + Q_i
\end{eqnarray}
with a 3-body phase space.
The tails of the lines extend to the energy limit imposed by the ground-state Q-value, and at that limit are sensitive to the modification of phase space caused by neutrino mass, just as in beta decay.    The existence of an electron-capture isotope, $^{163}$Ho, with a very low Q-value \cite{Kopp:2009yp} in the vicinity of 2.5 keV heightened the interest in this approach and a number of  experimental groups  explored the possibility with a variety of techniques \cite{Hartmann:1992jg,Yasumi:1982rb,Yasumi:1986hs,Yasumi:1994uu,Andersen:1982hh,Jonson:1982gr}.  

Advances in the art of microcalorimetry have spurred a resurgence of interest, as very high resolution spectra from large arrays of detectors become a possibility \cite{Meunier:1996ge,Meunier:1998pt,Gastaldo:2004ji,Ranitzsch:2012JLTP,NuMECS2014,Nucciotti:2014raa,Alpert:2014lfa}.  In a calorimetric experiment one is indifferent to the details of how the vacancy refills, whether by radiation or electron ejection, and records a spectrum of $E_c$, the ``visible energy''  ({\it i.e.}, that not carried away by the neutrino) converted to heat.   Three current experimental programs have been reported.  The ECHo project uses metallic magnetic calorimeters, which are composed of two Au carriers, one implanted with a paramagnetic ion (Er) and one with $^{163}$Ho nuclei.  The temperature change induced by a decay causes a magnetic flux change that can be read out via SQUID sensors.  Impressive 8.3-eV energy resolution has been achieved with these devices \cite{Ranitzsch:2012JLTP,Ranitzsch:2014kma}, and some recent data from ECHo are shown below.  It has recently been shown \cite{Prasai:2013bfa} that implanting Au carriers with both Er and Ho does not cause deleterious changes in the heat capacity at 110 mK.  
The NuMECS collaboration \cite{NuMECS2014} makes use of a different thermal sensor technology, Mo/Cu bilayer superconducting transition-edge sensors (TES).  Initial tests are under way with $^{55}$Fe.  The HOLMES collaboration has also recently selected the Mo/Cu TES technology \cite{Alpert:2014lfa}.

In order to make a convincing case about  the neutrino  mass  from this kind of experiment, it is important to understand what the spectrum would look like without it.  As there is no way to set the mass to zero experimentally, there is no recourse but to rely on theory.  

Ascribing a Breit-Wigner line shape to each vacancy and imposing a phase-space and energy-conservation envelope, \adr\ and Lusignoli calculate the spectrum to be expected \cite{DeRujula:1981ti,DeRujula1982429,DeRujula:2013jba}.  Expanding the neutrino flavor eigenstate in the mass basis, and following \cite{DeRujula:1981ti} and \cite{Nucciotti:2014raa}, one can obtain the spectrum in the following form:
\begin{eqnarray}
\frac{d\lambda_{\rm EC}}{dE_c}&=&\frac{G_F^2\cos^2{\theta_C}}{2\pi^3}(Q-E_c)\times \nonumber \\
&& \sum_i\left|U_{ei}\right|^2\left[(Q-E_c)^2-m_{\nu i}^2\right]^{1/2}\times \nonumber\\
&& \sum_j\beta_j^2C_j\left|M_{j0}\right|^2\frac{\Gamma_j}{4(E_c-E_j)^2+{\Gamma_j}^2}, \label{eq:spectrum}
\end{eqnarray}
where $G_F$ is the Fermi coupling constant and $\theta_C$ is the Cabibbo angle, $U$ is the neutrino mixing matrix and $m_{\nu i}$ is an eigenmass, $\beta_j$ is the amplitude of the electron wave function at the origin, $C_j$ is the nuclear shape factor, and $E_j$ and $\Gamma_j$ are the excitation energy and natural width of atomic configuration $j$.  The quantity $M_{j0}$ is an overlap (monopole) electronic matrix element between the ground state of the decaying atom and state $j$ of the daughter atom.  Exchange effects \cite{Faessler:2014xpa} and orbital occupancies are  here absorbed into $M_{j0}$.   In the specific case of $^{163}$Ho, the index $j$ runs over the 7 occupied orbitals from which capture can occur: (3s), (3p1/2), (4s), (4p1/2), (5s), (5p1/2), and (6s).

\begin{figure*}[ht]
 \vspace{.2in}
\centerline {
\includegraphics[width=\linewidth]{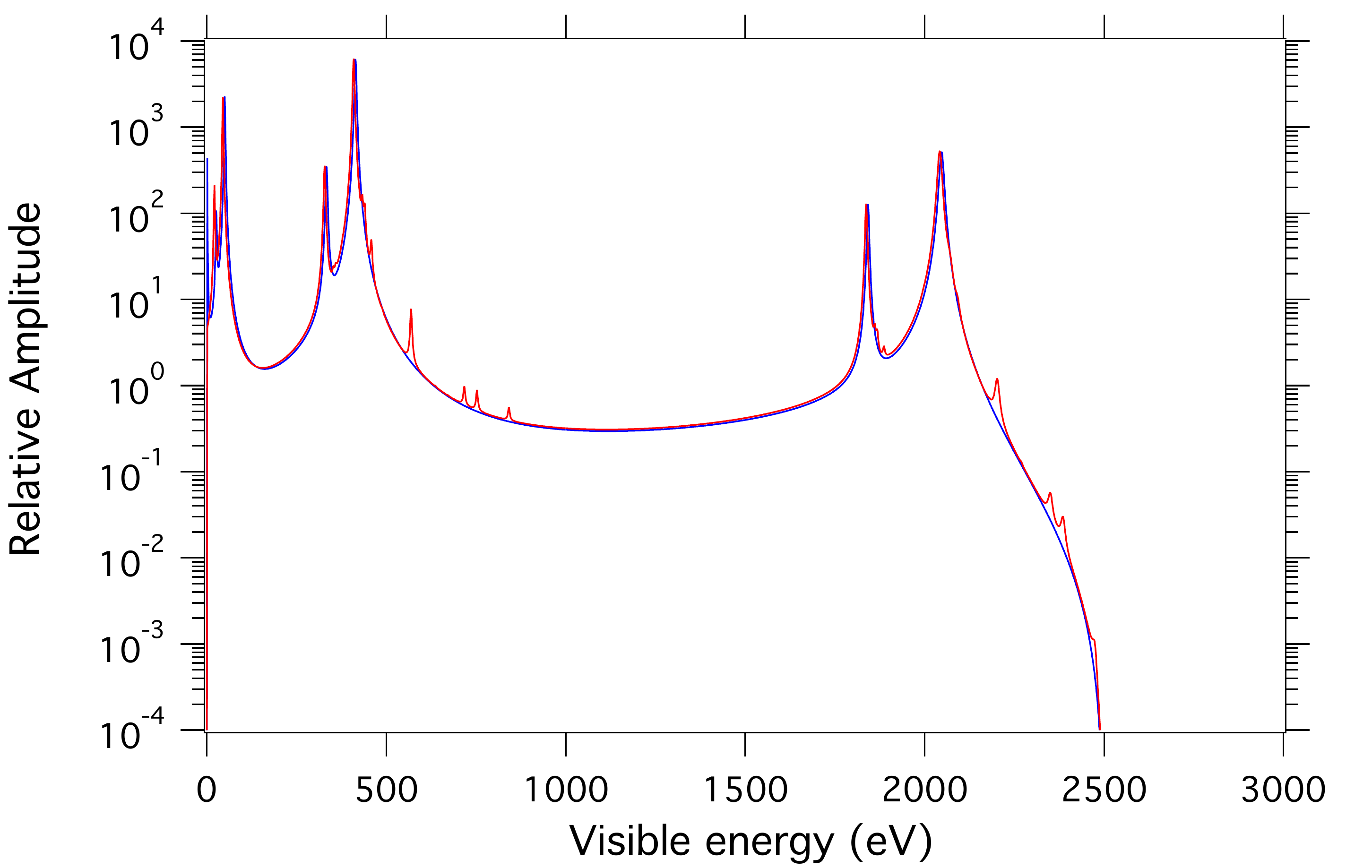}
}
\caption[]{(Color online) The visible energy in a calorimeter following electron capture in $^{163}$Ho.  The simpler spectrum (blue) is calculated in the customary single-vacancy approximation.  The more complex spectrum (red) includes configurations with 2 vacancies and an extra (4f7/2) electron.  The energies are  calculated using primary vacancy energies in $^{163}$Dy and secondary vacancy energies in $^{163}$Ho.  The shakeup probabilities for the satellite peaks are taken from calculations for Xe by Carlson and Nestor \cite{PhysRevA.8.2887}.   Primary vacancies  in the (6s)  shell and double vacancies in the (5sp) shell have not been considered.}
\vspace{.2in}
\label{fig:ecspectra}
\end{figure*}

Recent high-resolution calorimetric data \cite{Ranitzsch:2012JLTP,Ranitzsch:2014kma} confirm this expectation: there are sharp lines corresponding to the  energy released and thermalized as the vacancies refill.   The statistics are insufficient yet to reveal the wings in any detail, although the strong  (4s)$^{-1}$ NI line has shoulders  broader and with more structure than theory predicts.  Nevertheless, \adr\  argues that far from peaks the spectrum shape is determined only by phase space, and variation of the matrix element cannot be large enough to be relevant.  If so, the spectrum given in Eq.~\ref{eq:spectrum} can be used with confidence to predict the zero-mass shape near the endpoint and thereby derive experimental values for the neutrino masses.

However, while treating the capture in the simplified way described above with $j$ running over 7 single-particle orbitals is standard,  it is an approximation.   In what isotope is the vacancy formed, $^{163}$Ho or $^{163}$Dy?    An inner-shell electron has been absorbed suddenly in the nucleus, the nuclear charge has changed, and the index $j$ should range over the complete set of states energetically allowed in the 66-electron final-state Dy atom.  Included in that basis are many configurations of neutral Dy with two or more inner-shell vacancies and electrons in  bound but normally unoccupied valence levels, or in the continuum.    It might be thought that the probability of multiple vacancies must be very small.  On the contrary, for these rare earths, {\em all} final states have at least two atomic-orbit occupancies that are different from the ground-state configuration.  The ground states of Ho and Dy differ by a single (4f7/2) electron, but only  (s) and (p1/2) orbitals have sufficient amplitude at the origin for electron capture.  Hence the final state consists of at least an inner-shell vacancy and an extra (4f) electron.  While this particular circumstance modifies the  spectrum only slightly \footnote{It improves the accuracy of the calculated peak energies; compare the present results with ref. \cite{Ranitzsch:2014kma}}, more significant modifications result from additional vacancies in other shells.
 
\adr\  presents an estimate \cite{DeRujula:2013jba} that the rate for populating a Dy configuration with simultaneous (3s)$^{-1}$ and (4s)$^{-1}$ vacancies is $10^{-5}$ compared to a single (3s)$^{-1}$ vacancy and therefore negligible.  However, while the probabilities may decrease strongly with more complicated configurations, it is the total intensity near the endpoint that is relevant.  Population of the (3s)$^{-1}$(4s)$^{-1}$  configuration peaks where the single (3s)$^{-1}$  tail has become very weak and can even dominate the spectrum in that region.  The complex multi-vacancy configurations of neutral Dy include some that are `resonant'  in the sense defined by \adr: they have relatively narrow widths.  The vacancies refill by single-particle electromagnetic transitions, and we therefore assign them widths that are the same as the width of the primary (most deeply bound) vacancy, in the absence of experimental data.  The continuum shakeoff process is included with shakeup in the Carlson-Nestor (CN) theory \cite{PhysRevA.8.2887} adopted for the present analysis.  Shakeoff features are not as narrow as shakeup, but still give rise to enhancements at threshold with a higher-energy tail that falls off on a scale of tens of eV (see, for example, Ref.~\cite{PhysRevLett.67.2291}).  They are, therefore, also quite sharply defined spectral features.  When the atom is part of a solid, valence and continuum excitations of a still more complex nature become possible.

\section{Calculation}

This argument can be made more quantitative by considering the available configurations in this shakeup process and assigning  energies to each based on single-particle estimates.  The ground state of Dy I is ([Xe]4f$^{10}$6s$^2$).   The types of  excitations that can be present are restricted by the monopole selection rule, namely that the operator in the matrix element is the unit operator.  Restricting the space further to configurations that have only one or two vacancies and the extra (4f) electron, one can then construct a spectrum.  Table \ref{tab:coeffs2} lists the orbitals considered, and their binding energies.   The excitation energy of each configuration is assembled from the single-particle binding energies $E_b$ in Ho and Dy \cite{CRC}.  The innermost vacancy is taken to have a binding energy appropriate to Dy while less-bound shells are assigned binding energies appropriate to Ho, thereby allowing for the missing inner electron.  Both  are simplifying approximations that can be expected to lead to energy errors of a few eV.    The first three columns list the calculated intensity, visible energy, and width, respectively, of the primary and satellite features in the calorimetric spectrum.  The remaining columns identify an accessible configuration with an entry of $-1$ for a hole in a normally filled shell, and $1$ for a normally incomplete shell containing an extra electron.  

Single-vacancy capture probabilities are adopted from Lusignoli and Vignati  \cite{Lusignoli201111} which are in good agreement with the more recent results of Faessler {\em et al.} \cite{Faessler:2014xpa}, with the inclusion of overlap and exchange corrections.  For each primary vacancy, the relative populations of satellite shakeup configurations are taken from the calculations by Carlson and Nestor \cite{PhysRevA.8.2887}  for Xe.     The shakeup probabilities are thereby normalized to single-vacancy probabilities that include overlap and exchange corrections in Dy.   However, the CN vacancy probabilities are calculated in Xe (Z=54), not Dy (Z=66).  An approximate correction for this can be made by noting that the secondary vacancy probability is of order
\begin{eqnarray}
P&\simeq&1 - \braket{\phi^{\rm Dy}_{nlj}|\phi^{\rm Ho}_{nlj}}^{2(2j+1)} \label{eq:zscale}
\end{eqnarray}
and that, for Coulomb wavefunctions, the squared overlap integral  depends on Z as  $1-191/32Z^2$ \cite{DeRujula:1981ti}.   The CN probabilities for Xe are therefore rescaled by a factor $(54/66)^{2(2j+1)}$. 
 The resulting spectrum is shown in Fig.~\ref{fig:ecspectra}.  It is seen to be quite complex even in the relatively restricted space considered.  The appearance of a shakeup peak very close to the endpoint for the chosen Q-value, 2.5 keV, is accidental but underscores the difficulty in determining precisely the underlying spectrum, as would be required in order to make a definitive statement about neutrino mass from $^{163}$Ho electron capture.
 
\begin{figure}[ht]
 \vspace{.2in}
\centerline {
\includegraphics[width=\linewidth]{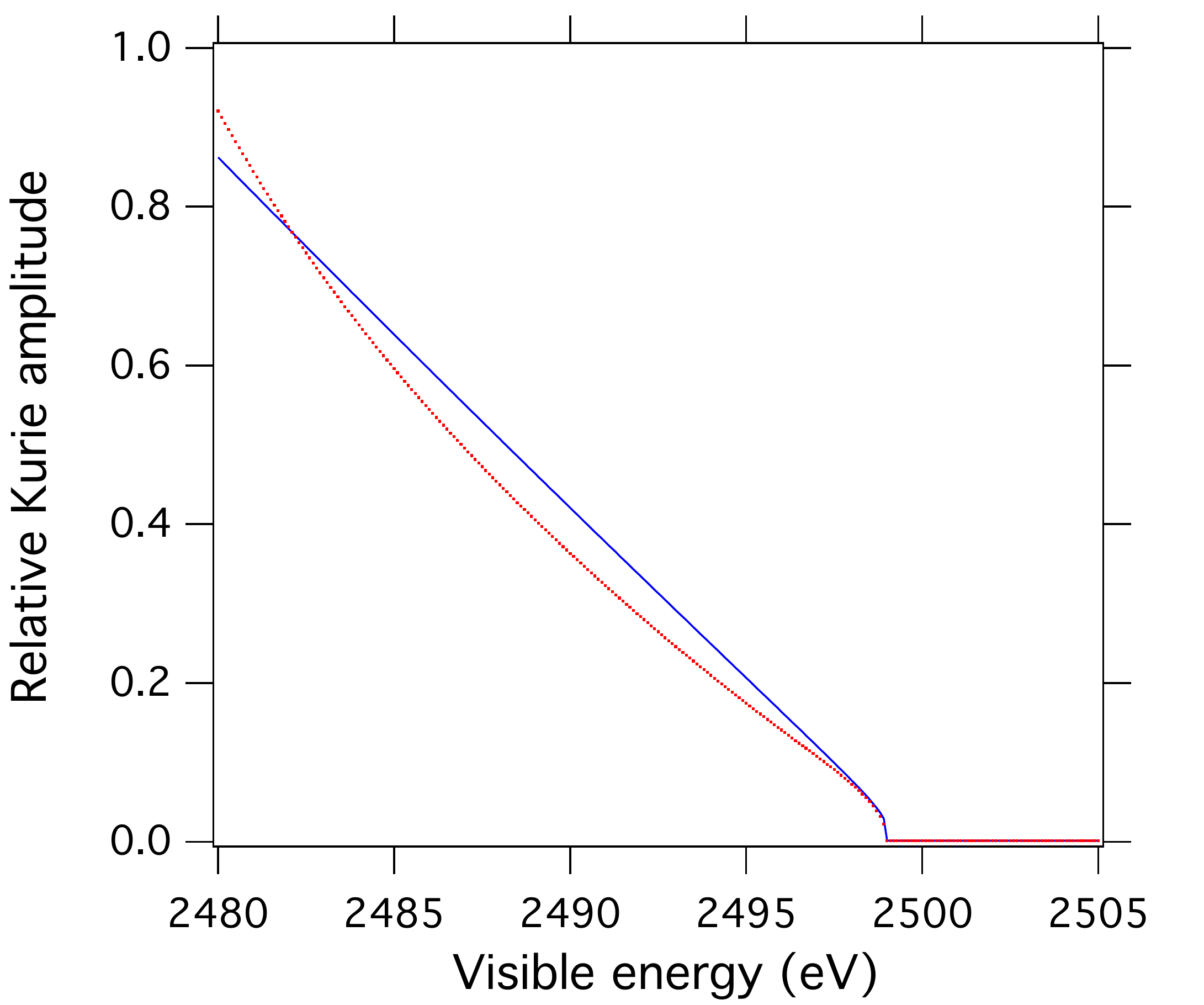}
}
\caption[]{(Color online) Expanded view of the region near the endpoint showing the spectrum without satellite structure included (blue, solid curve) and with it (red, dotted curve).  In both cases $Q=2500$ eV and $m_\nu^2 = 1$ eV$^2$.  The normalizations are arbitrarily chosen to bring out the differences in shape.  If the presence of the curvature in the spectrum near the endpoint were not known to an analyst, fitting to the standard spectral shape would produce erroneous results for $Q$ and $m_\nu^2$.}
\vspace{.2in}
\label{fig:kurie}
\end{figure}

 An expanded view of the region near the endpoint is shown in Fig.~\ref{fig:kurie}.    
The spectra are calculated with $Q=2500$ eV and $m_\nu^2 = 1$ eV$^2$ and are shown as Kurie plots, which present the square root of the spectral intensity.  The advantage of the Kurie representation is that the statistical weight of each point is the same, and that the `standard' spectrum is, to a very good approximation, a straight line except at the endpoint.   The curvature in the spectrum with satellite structure is produced by the tail of the nearby (3s)$^{-1}$(4s)$^{-1}$ double vacancy.  The procedure for extracting a neutrino mass from data involves fitting the spectrum to both $Q$ and $m_\nu^2$, because $Q$ is never well enough known to be fixed from independent data.  If one were unaware of the spectral distortion and attempted to fit the spectrum in this region with the standard shape, the result would be   $Q=2499$ eV and $m_\nu^2 = -16$ eV$^2$, very far from the correct values.  (Including background and instrumental resolution would change these numbers.)   Naturally, with the existence of these satellite peaks having been demonstrated in the present work, a better approach to fitting would be to include the curvature.  Unfortunately, however, the shapes of the satellite structures are unknown.  They are not simple Lorentzians as assumed for illustration, but a complex blend of overlapping bound and continuum line shapes from many final states.

\squeezetable
\begin{table*}
\caption{Energies, intensities, and occupation number differences for configurations populated in $^{163}$Ho electron capture. The single-particle binding energies  $E_b$ are in eV \cite{CRC}. \label{tab:coeffs2}}
\begin{ruledtabular}
\begin{tabular}{rrrrrrrrrrrrrrrrrr}
&$E_b$&&3s1/2&3p1/2&3p3/2&3d3/2&3d5/2&4s1/2&4p1/2&4p3/2&4d3/2&4d5/2&4f5/2&4f7/2&5s1/2&5p1/2&5p3/2\\

&Dy &&2047&1842&1676&1333&1292&414.2&333.5&293.2&153.6&153.6&8&4.3&49.9&26.3&26.3\\
&Ho &&2128&1923&1741&1392&1351&432.4&343.5&308.2&160&160&8.6&5.2&49.3&30.8&24.1\\
Rel.&&&&&&&&&&&&&&&&&\\
intens.&$E_c$&$\Gamma$&&&&&&&&&&&&&&&\\
\%&eV&eV&&&&&&&&&&&&&&&\\
100&2041.8&13.2&-1&&&&&&&&&&&1&&&\\
0.034&2474.2&13.2&-1&&&&&-1&&&&&&1&&&\\
0.049&2385.3&13.2&-1&&&&&&-1&&&&&1&&&\\
0.05&2350.0&13.2&-1&&&&&&&-1&&&&1&&&\\
0.211&2201.8&13.2&-1&&&&&&&&-1&&&1&&&\\
0.146&2201.8&13.2&-1&&&&&&&&&-1&&1&&&\\
0.609&2091.1&13.2&-1&&&&&&&&&&&1&-1&&\\
1.398&2072.6&13.2&-1&&&&&&&&&&&1&&-1&\\
1.468&2065.9&13.2&-1&&&&&&&&&&&1&&&-1\\
5.26&1836.8&6&&-1&&&&&&&&&&1&&&\\
0.002&2269.2&6&&-1&&&&-1&&&&&&1&&&\\
0.003&2180.3&6&&-1&&&&&-1&&&&&1&&&\\
0.003&2145.0&6&&-1&&&&&&-1&&&&1&&&\\
0.011&1996.8&6&&-1&&&&&&&-1&&&1&&&\\
0.008&1996.8&6&&-1&&&&&&&&-1&&1&&&\\
0.032&1886.1&6&&-1&&&&&&&&&&1&-1&&\\
0.074&1867.6&6&&-1&&&&&&&&&&1&&-1&\\
0.077&1860.9&6&&-1&&&&&&&&&&1&&&-1\\
23.29&409.0&5.4&&&&&&-1&&&&&&1&&&\\
0.001&841.4&5.4&&&&&&-2&&&&&&1&&&\\
0.002&752.5&5.4&&&&&&-1&-1&&&&&1&&&\\
0.002&717.2&5.4&&&&&&-1&&-1&&&&1&&&\\
0.015&569.0&5.4&&&&&&-1&&&-1&&&1&&&\\
0.011&569.0&5.4&&&&&&-1&&&&-1&&1&&&\\
0.114&458.3&5.4&&&&&&-1&&&&&&1&-1&&\\
0.282&439.8&5.4&&&&&&-1&&&&&&1&&-1&\\
0.302&433.1&5.4&&&&&&-1&&&&&&1&&&-1\\
1.19&328.3&5.3&&&&&&&-1&&&&&1&&&\\
0.00004&671.8&5.3&&&&&&&-2&&&&&1&&&\\
0.00009&636.5&5.3&&&&&&&-1&-1&&&&1&&&\\
0.00074&488.3&5.3&&&&&&&-1&&-1&&&1&&&\\
0.00052&488.3&5.3&&&&&&&-1&&&-1&&1&&&\\
0.00565&377.6&5.3&&&&&&&-1&&&&&1&-1&&\\
0.01371&359.1&5.3&&&&&&&-1&&&&&1&&-1&\\
0.01527&352.4&5.3&&&&&&&-1&&&&&1&&&-1\\
3.45&44.7&3&&&&&&&&&&&&1&-1&&\\
0.15&21.1&3&&&&&&&&&&&&1&&-1&\\
\end{tabular}
\end{ruledtabular}
\end{table*}

\begin{figure}[ht]
 \vspace{.2in}
\centerline {
\includegraphics[width=\linewidth]{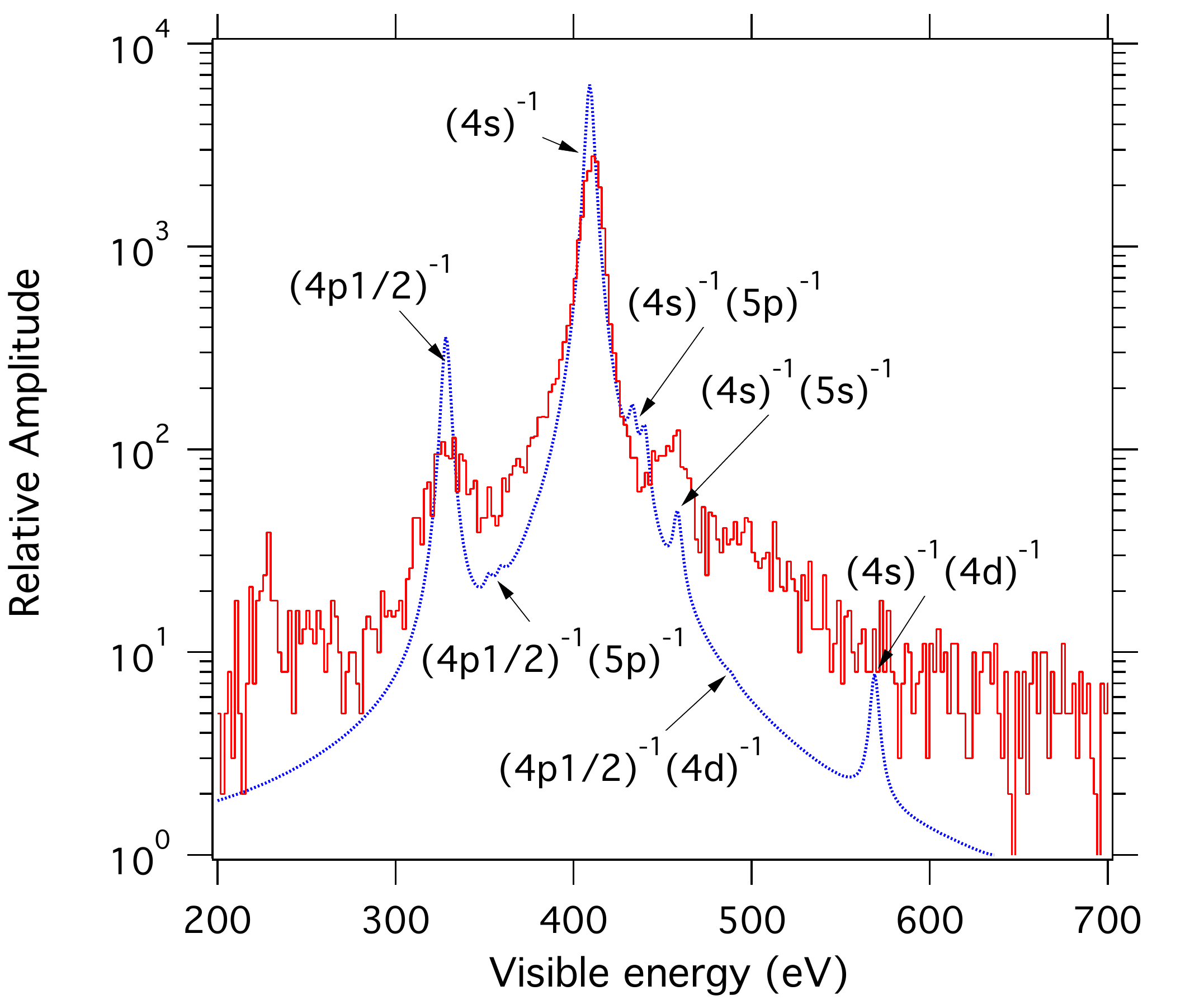}
}
\caption[]{(Color online) Expanded view of the vicinity of the NI line in the $^{163}$Ho spectrum recorded calorimetrically by Ranitzsch {\em et al.} \cite{Ranitzsch:2014kma} (solid line, red).  The calculated spectrum (blue dotted line) exhibits satellites on the high-energy side of the line, and the location of the (4s)$^{-1}$(5s)$^{-1}$ double vacancy at 458.3 eV corresponds to the observed satellite peak in the data.  However, the intensity predicted with CN theory is lower, and other satellites in the vicinity do not appear at the predicted intensity (see text).  The experimental resolution is given in  \cite{Ranitzsch:2014kma} as 8.3 eV; the theory is shown only with the assumed natural width of 5.4 eV. }
\vspace{.2in}
\label{fig:nline}
\end{figure}

There is evidence already in the data that shakeup satellites are present.  An unidentified peak is observed on the upper shoulder of the (4s)$^{-1}$ line in the high-resolution calorimetric study by Ranitzsch {\em et al.} \cite{Ranitzsch:2012JLTP,Ranitzsch:2014kma}.  In Fig.~\ref{fig:nline} this region is compared with the theoretical spectrum including shakeup and shakeoff satellites.  The unidentified satellite has the correct energy to be the (4s)$^{-1}$(5s)$^{-1}$ double vacancy line. 

The satellite spectrum presented here is merely indicative, rather than quantitative.   The shakeup and shakeoff calculations of Carlson and Nestor were carried out for  photoionization of Xe.   In photoionization, the electron is ejected from the atom, while in electron capture it is captured in the nucleus.   Thus in photoionization it is the outermost orbitals that are subjected to the largest change in effective charge, whereas in electron capture it is the innermost.   It is therefore not surprising that the CN calculation applied to electron capture would overestimate shakeup from higher-$j$ orbitals that do not have significant amplitude at the nucleus, nor, conversely, is it surprising that the sole double-vacancy state visible in the data so far is (4s)$^{-1}$(5s)$^{-1}$ at an intensity underestimated by the CN theory.   To deal with the `screening' (`antiscreening' would be more descriptive) provided by the inner-shell vacancy, \adr\  \cite{DeRujula:2013jba} evaluates an effective charge that screens the outer-shell vacancy.  For the  (3s)$^{-1}$(4s)$^{-1}$ double vacancy, the effective charge turns out to be almost unity (0.9649) and thus largely restores the lack of overlap that contributes to shakeup and shakeoff.  In this case,  \adr\  finds an intensity $1.08\times10^{-5}$, about 30 times smaller than the unscreened result using the CN calculation.  Indeed, the screening effect may be that large; with only approximate methods available at this time it is difficult to evaluate their accuracy.   Experimental input would be helpful,  if the identification of the (4s)$^{-1}$(5s)$^{-1}$ satellite can be confirmed and supplemented by observation of other satellites.

A less important mismatch is that for comparable excitations in Dy promotions into the (6s) shell are blocked,  reducing the phase space available compared to Xe.   Similarly, the increased binding of (5s5p) electrons in Dy compared to Xe can be expected to inhibit shakeup from those orbitals.     Many of the shortcomings could be addressed in a more advanced and specific theoretical treatment, but the conclusion that the spectrum is much more complex than has been assumed heretofore is one that does not depend on such refinements.

\section{Conclusion}

Electron capture in $^{163}$Ho measured calorimetrically offers a potentially attractive method for measuring neutrino mass. A quantitative understanding of the shape of the underlying spectrum with zero neutrino mass is essential for drawing  reliable conclusions  experimentally about the actual value of the mass.  An indication of the complexity of the spectrum has been presented.   Considering vacancy multiplicities of only 1 or 2, the spectrum is dense with line and edge features up to a Q-value of about 2.5 keV, but for larger Q-values up to about 3 keV the spectrum is featureless near the endpoint in this approximation.  This may offer an avenue for experiments if the Q-value is confirmed to be in the vicinity of 2.8 keV, as is indicated by recent studies \cite{Ranitzsch:2012JLTP,Ranitzsch:2014kma}.  A more detailed calculation could reveal if higher-order satellites or shakeoff features also populate that region.  Coherent interference between the tails of resonances and inner bremsstrahlung is another likely complication.   For the larger Q-values, the continuum phase space at the endpoint becomes so small \cite{Nucciotti:2014raa} that  the experimental measurement itself is very challenging.   Nevertheless, if the steadily improving experimental sensitivity is matched by new, more quantitative relativistic theoretical calculations, the  precise agreement between theory and experiment  necessary for a neutrino mass measurement may yet emerge.

The suggestion for Eq.~\ref{eq:zscale} by a referee is gratefully acknowledged.  This material is based upon work supported by the U.S. Department of Energy Office of Science, Office of Nuclear Physics under Award Number DE-FG02-97ER41020.

\bibliography{testbib}{}

\end{document}